\documentclass[12pt]{article}
\newcommand{\be}{\begin{equation}}
\newcommand{\ee}{\end{equation}}
\newcommand{\ba}{\begin{eqnarray}}
\newcommand{\ea}{\end{eqnarray}}

\def\LP{\displaystyle{\Biggl(}}
\def\RP{\displaystyle{\Biggr)}}

\newcommand{\dfrac}[2]{{\displaystyle{\frac{#1}{#2}}}}

\begin{document}

\title{\bf  Nonrelativistic Supersymmetry in
Noncommutative Space}
\author{ Gustavo~S.~Lozano$^a$\thanks{Associated with CONICET}\,,
Olivier~Piguet$^b$ \,,
Fidel~A.~Schaposnik$^c$\thanks{Associated with CICBA} \\ and
Lucas~Sourrouille$^a$
\\
{\normalsize\it $^a$Departamento de F\'\i sica, FCEyN, Universidad
de Buenos Aires}\\ {\normalsize\it Pab.1, Ciudad Universitaria,
Buenos Aires, Argentina}
\\
{\normalsize\it $^b$Universidade Federal do  Esp\'\i rito Santo,
UFES, Vit\'oria, ES, } {\normalsize\it  Brasil}
\\
 {\normalsize\it $^c$Departamento de F\'\i
sica, Facultad de Ciencias Exactas}\\
{\normalsize\it Universidad Nacional de La Plata, C.C. 67, 1900 La
Plata, Argentina}
\\
{\footnotesize  lozano@df.uba.ar, opiguet@yahoo.com,
fidelschaposnik@yahoo.co.uk, lsourrouille@yahoo.es} } \maketitle

\abstract{We analize a model of non relativistic matter in $2+1$
dimensional noncommutative space. The matter fields interact with
gauge fields whose dynamics is dictated by a Chern Simons term. We
show that it is possible to choose the coupling constants in such a
way that the model has and extended supersyemmetry and  Bogomolnyi
equations can be found.}

 \vspace{5mm}


In the past few years, field theories defined in noncommutative (NC)
space have received much attention mainly in connection with the
effective  low energy description of string theories
\cite{Douglas:2001ba}. For the particular case of $2+1$ dimensional
space, it has also been argued that Chern Simons theories in NC
space can be used as an effective description of the physics of the
Quantum Hall Effect \cite{Susskind:2001fb}, \cite{recent}.

Motivated by these facts, the extension to NC space of the Jackiw
and Pi model (JP) \cite{jackiwpi} of non relativistic matter
interacting with gauge fields whose dynamics is  governed by Chern
Simons fields was first considered in \cite{LMS}.

In ordinary space, this model, which is related to the physics of
the Aharonov-Bohm problem, provides a nontrivial example of a
gauge theory invariant under the action of the Galilean group
\cite{hagen}. Indeed, the space-time invariance group is larger as
the theory is also invariant under dilations and conformal
transformations, at least at the classical level. As it is the
case for many gauge theories, the scale invariance of the action
is broken by quantum corrections. Interestingly, the invariance is
recovered for a particular relation of coupling constants
\cite{bergmanlozano}.

In NC space, the model looses some of the symmetries present in
ordinary space. This is somewhat expected as non-commutativity
breaks explicitly the scale symmetry and the boost sector of the
Galilean group (for a detailed discussion see \cite{horvathy}).
Despite this fact, both versions of the model share some important
aspects, like the existence of BPS equations for a particular
relation of coupling constant \cite{jackiwpi},\cite{LMS},\cite{bak}.

In  ordinary space, the JP model admits and $N=2$ supersymmetric
extension \cite{leblanclozanomin}, providing another example of the
by know well established connection between BPS equations and
supersymmetry \cite{susy1}-\cite{susyn}.
This case is particularly interesting as it
provides a non trivial explicit realization of the graded Galilean
symmetry originally discussed in \cite{Puzalowski:1978rv}.

In this paper we shall examine possibility of building a
supersymmetric extension of the JP model in noncommuative space. We
shall show that indeed it is possible to do so precisely for the
same relation of coupling constants for which BPS equation exist.

\vspace{1cm}

We will  be interested in $d=2+1$ noncommutative space characterized
by the relations \be [x_1,x_2]=i \theta\;,\qquad [x_i,t]=0\;,
\label{1} \ee where $\theta$ is a real constant with dimension of
length squared. It will be convenient to introduce complex variables
$z$ and $\bar z$
\be
z = \frac{1}{\sqrt 2}(x^1 + i x^2) \; , \;\;\;\;\;\;
\bar z= \frac{1}{\sqrt{2}}(x^1 - i x^2)\;,
\label{2}
\ee
which can be  related to
 annihilation and creation operators  $\hat a$ and $\hat a^\dagger$
acting on a Fock space,
\be
\hat a = \frac{1}{\sqrt{\theta}} z \;, \;\;\;\;\;\;
\hat a^\dagger = \frac{1}{\sqrt{\theta}} \bar{z}\;,
\label{3}
\ee
so that (\ref{1}) becomes
\be
[\hat a,\hat a^\dagger] = 1\;.
\label{4}
\ee
In this way, through the action of $a^\dagger$ on the vacuum state
$|0\rangle$, eigenstates of the number operator
\be
\hat N = a^\dagger a
\label{number}
\ee
are generated.
With our conventions, derivatives in the Fock space are given by
\be
\partial_z = -\frac{1}{\sqrt \theta}
[\hat a^\dagger,~] \; , \;\;\;\;\;\; \partial_{\bar z} =
\frac{1}{\sqrt \theta} [\hat a,~] \;, \label{5} \ee
and integration on the noncommutative plane should be interpreted
as a trace

\be \int d^2x  \to 2\pi \theta {\rm Tr}  \;. \ee

We are interested in the  model of nonrelativistic matter interacting with gauge fields whose
dynamics is governed by the Chern-Simons term that has been considered in \cite{LMS}.
The action associated to this model can be written as
\begin{eqnarray}
S=S_{cs} + \int \left(   i \phi^\dagger D_t \phi - \frac{1}{2m} ( D_i\phi
)^\dagger  D_i\phi
 + \lambda_{1} (\phi^\dagger  \phi \phi^\dagger  \phi) \right)d^3 x\;.
\label{action1}
\end{eqnarray}
Here, $S_{cs}$ is the Chern Simon action\footnote{ Throughout this
paper we will be using the   notation  $V^{\pm}=V^1\pm iV^2$ where
$V^1$ and $V^2$ are the components of any vector $V$ in the plane.}
 \be S_{cs}=
  \int d^3x \left( - \frac{\kappa}{4ci}[(\partial_t A)_+ A_-
-(\partial_t A)_- A_+]- A_{0} \kappa B_{12} \right)\;,
\ee
where
\begin{eqnarray}B_{12}=\frac{1}{2i}(\partial_-A_+-\partial_+A_-)-
\frac{e}{2c}[A_-,A_+] \;,
\end{eqnarray}
and $\phi$ denotes a complex {\it bosonic} field. The interaction
with the gauge fields is introduced via the covariant derivatives
\be
D_t\phi=\partial_t \phi +i e A_o \phi \;,\,\,\,\,\,
D_i\phi=\partial_i \phi -ie A_i \phi \;.
\label{fund}
\ee
The model is invariant
under gauge  transformations,
\be
\phi'=U\phi \;\; \psi'=U\psi
\;\;\; A_i=U^{-1}A_iU-\frac{i}{e}\partial_i U U^{-1}\;.
\label{fundi}
\ee
 The model
described by the action (\ref{action1}) is the realization in NC
space of the one originally discussed in
 \cite{jackiwpi}. Notice that in the NC case one needs to choose
 a particular
 ordering in the covariant derivative. For definiteness,
  we will be working with the
 ``fundamental'' representation (\ref{fund})-(\ref{fundi}) but the other
 cases can be handled
 similarly.

In order to explore the possibility of building a supersymmetric
extension of the model, we enlarge the field content of the theory
and include a nonrelativistic fermion $\psi$. The action of the model
 then becomes,
\begin{eqnarray}
S&=& S_{cs}+\int d^3 x \left( i \phi^\dagger  D_t \phi + i \psi^\dagger
D_t \psi - \frac{1}{2m} ( D_i\phi )^\dagger  D_i\phi -\frac{1}{2m}
( D_i\psi )^\dagger D_i\psi  \right.\nonumber \\
&+&  \frac{e}{2mc} \psi^\dagger  B_{12} \psi  + \left. \lambda_{1}
(\phi^\dagger  \phi \phi^\dagger  \phi) + \lambda_{2}
(\phi^\dagger  \phi  \psi^\dagger  \psi )+ \lambda_{3} (\phi
\phi^\dagger  \psi \psi^\dagger ) +\lambda_{4} (\psi^\dagger \psi
\psi^\dagger \psi)\right)  \nonumber\\
\label{action1-1}
\end{eqnarray}
%
%
This action is the simplest generalization to noncommutative space
of the one studied in Ref.\cite{leblanclozanomin}. Notice that terms
of the potential proportional to $\lambda_2$ and $\lambda_3$ would
be equivalent to each other in ordinary space and that the term
proportional to $\lambda_4$ would be identically zero due to the
anticommuting character of the  fermion fields. We have chosen a
particular sign for the Pauli interaction (corresponding to a "down"
spinor) which coincides with the one in \cite{leblanclozanomin}.

In a nonrelativistic setting the action of bosons and fermions are
almost identical if it were not for the presence of the Pauli term
for fermions. As shown below, the supersymmetry
variation of this last term is
easily compensated by the Chern Simons term.

 We will calculate the
variation of the action under the following   supersymmetry
transformation.
\begin{eqnarray}
\delta_{1} \phi &=& \sqrt{2m} \eta_{1}^\dagger  \psi  \;, \;\;\;
\;\;\;
\delta_{1} \psi= - \sqrt{2m} \eta_{1} \phi\;, \nonumber \\
\delta_{1} A &=& 0 \;, \;\;\; \;\;\;
\qquad\quad\ \delta_{1} A^{0}=
\frac{e}{\sqrt{2m}c \kappa} ( \eta_{1} \phi \psi^\dagger  -
\eta^\dagger _{1} \psi \phi^\dagger  ) \;. \label{var1}
\end{eqnarray}
It is trivial to show that,
\begin{eqnarray}
& &
 \delta_1\int\left( -\frac{\kappa}{4ic}((\partial_t A)_+ A_- -
(\partial_t A)_- A_+)+ i\phi^\dagger
\partial_t \phi + i
\psi^\dagger  \partial_t \psi \right) d^3 x=0 \;,\nonumber \\
& &
 \delta_1\int \left( \frac{1}{2m}(D\phi)^\dagger (D\phi) +
(D\psi)^\dagger (D\psi) \right) d^3 x = 0 \;, \nonumber \\
& & \delta_1
\int\left( -A_{0} \kappa B_{12} + \frac{e}{2mc}\psi^\dagger B_{12}\psi
\right) d^3 x =0\;.
\end{eqnarray}
Then the variation of the action reduces to
\begin{eqnarray}
 \delta_1 S&=&
\sqrt{2m}\int d^3
x \left(\eta_1[(-\frac{e^2}{2mc\kappa}-2\lambda_1-\lambda_3+\lambda_2)\phi^\dagger \phi\psi^\dagger \phi
\right.\nonumber\\
&& + \left. (\frac{e^2}{2mc\kappa}-\lambda_3
+2\lambda_4-\lambda_2)\psi^\dagger \phi\psi^\dagger \psi] + h.c.\right)\;.
 \label{var1n}
\end{eqnarray}
So the transformations in   eq.(\ref{var1}) correspond to a symmetry
of the action if the following relations are satisfied,
\begin{eqnarray}
\frac{e^2}{2mc\kappa} + 2\lambda_1 + \lambda_3-\lambda_2 = 0\;,
\nonumber \\
\frac{e^2}{2mc\kappa}-\lambda_3 +2\lambda_4-\lambda_2 = 0 \;.
\label{ac1}
\end{eqnarray}
While the first condition coincides with the one arising in ordinary
space, the second  one is peculiar to noncommutative space.
It originates from the second term in Eq(\ref{var1}) which is
automatically zero in ordinary space due to the   Grassman character
of $\psi$ and $\psi^\dagger $.

The fact that is less evident is the existence of a second
supersymmetry. Let us  examine the variation of the action under the
following transformation
\begin{equation}\begin{array}{l}
\begin{array}{ll}
\delta_{2} \phi = \dfrac{i}{\sqrt{2m}} \eta_{2}^\dagger  D_{+} \psi
\; ,  \qquad\qquad
&\delta_{2} \psi = - \dfrac{i}{\sqrt{2m}} \eta_{2} D_{-} \phi\;,
\nonumber \\[3mm]
\delta_{2} A^{+} = \dfrac{2e}{\sqrt{2m} \kappa} \eta_{2} \phi \psi^\dagger
\;,  \;\;\;
&\delta_{2} A^{-}= - \dfrac{2e}{\sqrt{2m}} \eta^\dagger _{2}
\psi \phi^\dagger \;, \nonumber \\
\end{array}\\[3mm]
\ \, \delta_{2} A^{0}   = \dfrac{ie}{( 2m )^{\frac{3}{2}} c \kappa}
( \eta_{2} \phi ( D_{+} \psi )^\dagger  + \eta^\dagger _{2}
( D_{+} \psi ) \phi^\dagger  ) \;.
\label{tiene}
\end{array}\end{equation}
Again it is easy to show the invariance of the kinetic terms,
\begin{eqnarray}
\delta\int\left( -\frac{\kappa}{4ic}((\partial_t A)_+ A_- -
(\partial_t A)_- A_+)+ i\phi^\dagger
\partial_t \phi + i
\psi^\dagger  \partial_t \psi \right) d^3 x = 0\ .
\end{eqnarray}

After some algebra, one can see that the variation of the remaining part of the action  can be written as
\begin{equation}
\delta_2 S= \delta_2^{(1)} S + \delta_2^{(3)} S\;.
\end{equation}
where $ \delta_2^{(1)} S$ is linear in fermion fields and $\delta_2^{(3)} S $
is cubic:
\begin{eqnarray}
\delta_2^{(1)} S&=&\frac{i}{\sqrt{2m}} \eta_2 \int d^3x \LP
\psi^\dagger  \left( \frac{e^2}{2mc\kappa} 2\phi(D_+\phi)^\dagger \phi
+\phi\phi^\dagger (D_-\phi) \right) \nonumber\\
& &
-2\lambda_1 (\phi(D_+\phi)^\dagger \phi +
\phi\phi^\dagger (D_-\phi) + (D_-\phi)\phi^\dagger \phi) + \nonumber\\
& &
\lambda_2(D_-\phi)\phi^\dagger \phi
-\lambda_3 (D_-\phi)\phi\phi^\dagger   \RP +
  h.c.\;,
\label{eq2}
\end{eqnarray}

\begin{eqnarray}
\delta_2^{(3)} S &=&\frac{i}{\sqrt{2m}} \eta_2 \int \LP\ d^3x \psi^\dagger
\left( \frac{e^2}{2mc\kappa}( \phi (D_+\psi)^\dagger \psi+
2 \psi(D_+\psi)^\dagger \phi \right) + \nonumber \\
& &
\lambda_2\psi (D_+\psi)^\dagger \phi + \lambda_3 \phi(D_+\psi)^\dagger \psi
-2\lambda_4(\phi(D_+\psi)^\dagger \psi +
\nonumber \\
&& \phi\psi^\dagger (D_-\psi) +\psi (D_+\psi)^\dagger \phi) ) \RP +
h.c. \label{eq3}
\end{eqnarray}
From eq.(\ref{eq2}), we obtain
\begin{eqnarray}
\frac{e^2 }{mc\kappa}-2\lambda_1 &=& 0\;, \nonumber \\
\frac{e^2}{2mc\kappa}-2\lambda_1 - \lambda_3 &=&0\;, \nonumber \\
-2\lambda_1 + \lambda_2&=&0\;,
\end{eqnarray}
while from eq.(\ref{eq3}) we get,
\begin{eqnarray}
\frac{e^2}{2mc\kappa}+ \lambda_3 - 2\lambda_4 &=&0 \;,\nonumber \\
\lambda_4 & =&0 \;,\nonumber \\
-\frac{e^2 }{mc\kappa}+\lambda_2 -2\lambda_4 & =&0\;.
\end{eqnarray}
The solution for this system is
 \be
 \lambda_1 =
\frac{e^2}{2mc\kappa} \; , \;\;\;\; \lambda_2 =
\frac{e^2}{mc\kappa}\; , \;\;\;\; \lambda_3 =
-\frac{e^2}{2mc\kappa} \; , \;\;\;\; \lambda_4 = 0\;, \label{cc}
 \ee
 which also   satisfies  the system   of eqs.(\ref{ac1}).

The results in ordinary space, namely \be \frac{e^2}{2mc\kappa} +
2\lambda_{1c} - \lambda_{2c}=0 \; , \;\;\;\;
\lambda_{1c}=\frac{e^2}{2mc\kappa} \;. \ee can be recovered by
noticing that the potential in ordinary space becomes,
\begin{eqnarray}
V &=& \lambda_1\phi^\dagger \phi\phi^\dagger \phi +
(\lambda_2-\lambda_3)\phi\phi^\dagger \psi\psi^\dagger\;,
\end{eqnarray}
so that  $\lambda_{2c} = \lambda_2 - \lambda_3$ can be identified.
After this identification, the ordinary space results follow.


The relation of coupling constants which make the model $N=2$
supersymmetric is connected to the BPS point. This can be most easily
seen by writing the Hamiltonian of the model as,
 \be H=\int d^2x
\left(\frac{1}{2m} (D_i\phi)^\dagger  D_i \phi +\frac{1}{2m}
(D_i\psi)^\dagger D_i \psi -\frac{e}{2m} \psi^\dagger  B \psi
+V(\phi,\psi) \right)\;,\ee where \be V[\phi,\psi]=-\lambda_1
\phi^\dagger  \phi \phi^\dagger  \phi-\lambda_2 \phi^\dagger
\phi\psi^\dagger \psi -\lambda_3\phi \phi^\dagger \psi \psi^\dagger
-\lambda_4 \psi^\dagger \psi \psi^\dagger \psi\;. \ee and
  \be B=-\frac{e}{\kappa}(\phi
\phi^\dagger -\psi \psi^\dagger )\;, \label{gl} \ee
Using the identity,
\be
(D_i\phi)^\dagger
D_i\phi=(D_{\pm}\phi)^\dagger  D_{\pm}\phi \pm e \phi^\dagger B\phi \;,
\label{bogo}
\ee
the Hamiltonian  can be re-written (up to surface terms) as
\begin{eqnarray}
H&=& \int d^2x \left(
\frac{1}{2m}  (D_{\pm}\phi)^\dagger  D_{\pm} \phi +\frac{1}{2m}
(D_{\pm}\psi)^\dagger  D_{\pm} \psi -(\lambda_1 \pm
\frac{e^2}{2m\kappa})\phi^\dagger  \phi \phi^\dagger  \phi \right.
\nonumber \\
& & -\lambda_2 \phi^\dagger \phi\psi^\dagger \psi +(-\lambda_3\pm
\frac{e^2}{m\kappa}-\frac{e^2}{2m\kappa})\phi \phi^\dagger \psi \psi^\dagger  +
\nonumber \\
& & \left. (-\lambda_4 \pm \frac{e^2}{2m\kappa}
-\frac{e^2}{2m\kappa})\psi^\dagger \psi \psi^\dagger \psi \right)\;.
\label{vaca}
\end{eqnarray}
In the bosonic sector of the theory, we then have,
 \be H= \int d^2x
\left( \frac{1}{2m}  (D_{\pm}\phi)^\dagger  D_{\pm} \phi-(\lambda_1
\pm \frac{e^2}{2m\kappa})\phi^\dagger  \phi \phi^\dagger  \phi
\right) \label{boso} \ee
 Thus, taking the lower sign,
 \be H= \int
d^2x \frac{1}{2m}  (D_{\pm}\phi)^\dagger  D_{\pm} \phi
\ee leads
to the BPS equation, \be D_-\phi=0 \label{boso2}
\ee

The particular
choice of sign in eq.(\ref{boso}) which leads to the
``anti-selfdual'' eq.(\ref{boso2}) is
related to our earlier  choice of sign for the Pauli
interaction. Changing this sign and redefining  the
susy transformations  accordingly leads to the
``self-dual'' BPS equation.

 In ordinary space, the choice that
makes the model $N=2$ supersymmetry invariant allows to write the
{\it full} Hamiltonian (i.e {\it bosons + fermions}) as a sum of
squares~\cite{leblanclozanomin}, just the first two terms in
(\ref{vaca}). Nevertheless, in NC space, one obtains,
\begin{eqnarray}
H&=&\int d^2x \left(\frac{1}{2m} (D_-\phi)^\dagger  D_ -\phi
+\frac{1}{2m} (D_-\psi)^\dagger D_- \psi   \right. \nonumber \\& &
 \left. -\frac{e^2}{m\kappa}
\phi^\dagger \phi\psi^\dagger \psi -\frac{e^2}{m\kappa}\phi \phi^\dagger \psi \psi^\dagger
- \frac{e^2}{m\kappa} \psi^\dagger  \psi \psi^\dagger  \psi
\right)\;.
\end{eqnarray}
The extra terms cancels in the commutative limit.

In order to write down the supersymmetry algebra, we start by
defining Poisson brackets. Calling $F, G$ the supersymmetry
charges or their Hermitian conjugates, we have
\be \left\{F,G\right\} = i \sum_j \int d^2x \left(\frac{\delta
F}{\delta \Omega_j(x,t)} \frac{\delta G}{\delta \Pi_j(x,t)} -
(-1)^{f_j} \frac{\delta F}{\delta \Pi_j(x,t)} \frac{\delta G
}{\delta \Omega_j(x,t)} \right)\;, \label{34}
\ee with
 \be
\Omega = \left(\phi,\psi,\sqrt{\frac \kappa 2}A_+\right) \; ,
\;\;\;
 \Pi = \left(i \phi^\dagger ,i \psi^\dagger ,i\sqrt{\frac \kappa 2}A_-\right)
  \; , \;\;\; f=(0,1,0)\;.
\ee
 Using Noether's theorem, the supersymmetric transformations
(\ref{var1}) and (\ref{tiene}) lead to the charges $Q_1$ and $Q_2$,
\be Q_1 = i\sqrt{2m} \int d^2x \phi^\dagger \psi \;,\label{car1} \ee
 \be
Q_2= \frac{1}{\sqrt{2m}}\int d^2x \left(D_{-}\phi\right) ^{\dagger} \psi\;.
\ee
Using (\ref{34}) one gets for the $Q_1$ bracket
\be
 \left\{Q_1,Q_1^\dagger \right\}= 2 m\int d^2x (\phi^\dagger \phi
 + \psi^\dagger \psi) \equiv 2M\;,
\ee
where we have introduced the total mass $M$.

Concerning the $Q_2$ bracket, one has
\be
\left\{Q_2,Q_2^\dagger \right\}=
\frac{1}{2m} \int  d^2z\left(
 \left(D_+ \psi\right)^{\dagger}(z) D_+ \psi(z) +
\left(D_-
\phi\right)^{\dagger}(z)  D_-\phi (z)
-\frac{2e^2}{\kappa} \phi^\dagger \psi^\dagger \psi \phi
\right)\;. \label{last}
\ee
Now,  identity (\ref{bogo}) allows us to rewrite eq.(\ref{last}) in the form
\begin{eqnarray}
\left\{Q_2,Q_2^\dagger \right\} &=&
\frac{1}{2m} \int  d^2z\left(\left(D_i \psi\right)^{\dagger}(z) D_i \psi(z) +
\left(D_i
\phi\right)^{\dagger}(z)  D_i\phi (z) -\frac{2e^2}{\kappa} \phi^\dagger \psi^\dagger \psi \phi
\right. \nonumber\\
&& \left.  + e \phi^\dagger  B \phi -e \psi^\dagger  B \psi
\vphantom{\left(D_i \psi\right)^{\dagger}}\vphantom{\frac{2e^2}{\kappa}}
\right) \;,\label{last2}
\end{eqnarray}
which, after using the Gauss law (eq.(\ref{gl})) becomes \be
\left\{Q_2,Q_2^\dagger \right\}= H\;. \ee
Finally, the only nonvanishing remaining bracket gives
\be
\left\{Q_1,Q_2^\dagger \right\} = \frac{1}{2i} \int d^2x
(\phi^\dagger D_-\phi - (D_-\phi)^\dagger \phi + \psi^\dagger D_-\psi -
(D_-\psi)^\dagger \psi) = P_- \;,
\ee
being $P_i$ the momentum.

As in the ordinary space case notice that the configurations of
fields $(\phi,\psi)$ such that \be D_-\phi=0 \;\;\;\; \psi=0 \ee is
left invariant by the supersymmetric transformation Eqs
(\ref{tiene}).

\vspace{1cm}

In summary, we have  been able to show that  a $N=2$
supersymmetric extension of  model for non relativistic matter
interacting with Chern Simons gauge fields can be built. This was
achieved for a particular relation of coupling constants given in
eq.(\ref{cc}), which in the commutative space limit $\theta=0$
reduces to the ordinary space result \cite{leblanclozanomin}. In
NC space, even for a U(1) gauge group  one has different
possibilities to couple gauge and matter fields according to the
choice of the covariant derivative (in the fundamental,
 the anti-fundamental or
the adjoint representation) and this can in principle lead to
different sets of BPS equations and their corresponding solutions.
Also, since the NC extension of the Jackiw-Pi model
\cite{jackiwpi} could be of relevance in connection with the
physics of the quantum Hall effect,
  the case of nonrelativistic matter interacting
with {\it non-Abelian} gauge fields would be
of much interest. In this respect, we believe that our results can be
extended to the case of a $U(N)$ group group as it
was done in ordinary space for the $SU(N)$ case
\cite{jackiwpiback}. We expect to report on these issues in the future.

\vspace{1cm}

We acknwoledge financial support from a CAPES-SECyT grant
$\#59/03$. F.A.S. is partially supported by a CICBA grant.
G.S.L
thanks the ICTP, Trieste, were part of this work has been done and
Carlos Nu\~nez for interesting discussions.


\end{document}